# Content Reuse and Interest Sharing in Tagging Communities

**Elizeu Santos-Neto[*], Matei Ripeanu[*], Adriana Iamnitchi[+]**

University of British Columbia, Electrical and Computer Engineering Department[*]
2332 Main Mall, Vancouver, BC, Canada V5
University of South Florida, Department of Computer Science and Engineering[+]
4202 E. Fowler Ave, Tampa, FL
{elizeus,matei}@ece.ubc.ca, anda@cse.usf.edu

**Abstract**

Tagging communities represent a subclass of a broader class of user-generated content-sharing online communities. In such communities users introduce and *tag* content for later use. Although recent studies advocate and attempt to harness social knowledge in this context by exploiting collaboration among users, little research has been done to quantify the current level of user collaboration in these communities. This paper introduces two metrics to quantify the level of collaboration: *content reuse* and *shared interest*. Using these two metrics, this paper shows that the current level of collaboration in CiteULike and Connotea is consistently low, which significantly limits the potential of harnessing the social knowledge in communities. This study also discusses implications of these findings in the context of recommendation and reputation systems

## Introduction

Several studies highlight the advantages of the collaborative nature of online communities like tagging communities (Golder and Huberman 2006), (Cattuto et al. 2007) (Hotho et al. 2006). In particular, (Golder and Huberman 2006) and (Iverson 2007) argue that collaborative tagging can turn into a powerful collaborative content management tool, if built upon personally driven incentives. The underlying assumption, shared by a number of other studies (Wu et al. 2006), (Yanbe et al. 2007), is that the collaborative nature of these communities makes the harvesting of social knowledge possible. An additional assumption is that such social knowledge can be exploited to increase the 'utility' delivered to users by building new applications or improving the services that support these communities (e.g., personalized search, recommendation systems, reputation, content distribution).

However, while the number of content sharing communities that incorporates tagging features continues to grow, existing research rarely discusses whether the level of collaboration in existing tagging communities supports these assumptions.

This paper aims to fulfill exactly this gap: *it evaluates the level of collaboration in existing tagging communities and provides a preliminary assessment of the conditions for efficiently harnessing the social knowledge generated in collaborative online communities.*

To quantify the conditions we are investigating, we propose two metrics that estimate the level of user collaboration in tagging communities: *content reuse* and *shared user interest*. In short, content reuse adopts an item-centric view and quantifies the dynamicity of a community in terms of the set of consumed items. The shared interest metric adopts a user-centric view and attempts to quantify the overlap between users' activities such as content sharing.

It is important to note that content reuse and shared interest interplay. Although a community may exhibit high levels of content reuse, its members can still have disjoint interests. This happens, for example, when each item is of particular interest for a single user, who keeps updating the set of tags on this particular item. On the other hand, the presence of shared user interest does imply a minimum level of content reuse, since shared user interest captures the collaboration among users with respect the tagging activity over a set of items.

This paper presents an evaluation of two popular collaborative tagging communities (CiteULike – http://citeulike.org and Connotea – http://connotea.org) in the light of the above metrics. Our analysis shows that *these communities exhibit consistently low levels of content reuse and shared interest*. Additionally we introduce a new data structure, the interest-sharing graph, to analyze the structure of these communities based on user tagging activity. In brief, the contributions of this paper are:

- This study identifies user collaboration as a condition to harness social knowledge in collaborative tagging communities, and proposes two metrics: content reuse and shared-user interest to evaluate it.
- This study characterizes two popular online communities in the light of the defined metrics. As opposed to widespread assumptions, we discover low levels of collaboration.
- Finally, we discuss the impact of these findings on the design of collaborative tagging systems and the challenges involved in building recommendation and reputation systems in these online communities

The rest of this paper is organized as follows. The next section defines collaborative tagging and presents related work on harnessing social knowledge. Section 3 describes the data sets. Section 4 analyzes the level of collaboration and its structure in CiteULike and Connotea. The paper concludes with a discussion on the impact of these findings.

## Background and Motivation

This section is divided into two parts. First, it introduces collaborative tagging and the terminology used in this paper. Then, it briefly reviews previous work related to harnessing social knowledge in these communities.

### Tagging Communities

Tagging allows users to attach strings from an uncontrolled vocabulary to data items such as URLs, pictures, or documents. Generally, collaborative tagging communities are organized around a web site that supports the mechanics of tagging together with additional functionality: adding new data items, search based on tags and/or on item content, authentication and authorization, and reputation.

The action of tagging an item performed by a user is referred to as a *tag assignment*. Additionally, adding an item to the community is referred to as *item posting*.

The collaborative nature of tagging relies on the fact that users potentially share interests, and, consequently, they post similar items and assign similar tags to them. Although users are self-centered while tagging (Golder and Huberman 2007), if tags are visible, this may facilitate the job of other users in finding content of interest.

### Harnessing Tagging Communities

A number of studies propose techniques to extract social knowledge in tagging communities.

Wu et al. (Wu et al. 2006) argue that, to turn collaborative tagging communities into efficient knowledge management tools, these communities must feature three mechanisms: community identification, user/document recommendation, and ontology generation. The proposed mechanisms are not evaluated quantitatively or qualitatively with respect to their utility and impact. We note, however, that the efficiency of these three mechanisms depends on level of collaboration in the community. In particular, if users do not reuse content or do not share interest, recommendation systems are likely to have little information to harness. Hence, it is unlikely to produce useful hints.

Yanbe et al. (Yanbe et al. 2007) suggest the use of content generated in tagging communities to improve the quality of web searches. In particular, they use del.icio.us (http://del.icio.us) to improve PageRank (Brin and Page 1998) rankings. Their hypothesis is that link-based ranking strategies, such as PageRank, do not capture fresh pages, even though they are relevant. Therefore, Yanbe et al. propose to use URL popularity in del.icio.us and combine it with the PageRank rankings to produce more relevant hints. They show that this approach improves item freshness at the top of the ranking while maintaining high relevance. The higher the content reuse and the shared interest in a community, the more efficient the above approach will be.

Halpin et al. (Halpin et al. 2006) propose to extract ontologies based on collaborative tagging activity. They analyze del.icio.us to determine how patterns can be used for such purpose. First, they show the tag frequency distribution of the 100 most tagged URLs can be approximated as a power law. Halpin et al. reason that the power-law popularity distribution in tag frequency is a strong indication that users have collaboratively achieved a consensus on which tag best represents a particular item. If users' tag assignments do not converge, due to a low level of shared interest, inferring ontologies may not be as efficient as when the community exhibits high levels of shared user interest.

*Summary*: all the techniques above have in common that they assume a high level of collaboration in the tagging communities they harness. In the next section, we cast doubt on these assumptions by quantitatively analyzing two relevant tagging communities.

## The Data Sets

We evaluate two online tagging communities: CiteULike and Connotea. They are designed as personal content management tools with collaborative features such as tagging and comments.

The data sets consist of all tagging activity since the creation of each community, more than two years of user activity for each. We obtained the CiteULike dataset directly from www.CiteULike.org website which provides logs of past tagging activity. For Connotea, we built a crawler that leverage Connotea's API to collect all data available since December 2004.

|                  | CiteULike         | Connotea          |
|------------------|-------------------|-------------------|
| **Activity period** | 11/2004—08/2007 | 12/2004—07/2007 |
| **# Users**      | 21,980            | 10,667            |
| **# Items**      | 625,048           | 267,150           |
| **# Tags**       | 188,301           | 110,276           |
| **# Assignments**| 3,342,694         | 891,005           |

Table 1: The data sets evaluated.

Table 1 presents the characteristics of the data sets analyzed. It is worth highlighting that we only have access to traces of explicit content use (i.e., tag assignments and item postings). An entry in the activity trace means that a user assigned a particular tag to one item, at a particular timestamp. The analysis of implicit content usage traces (i.e., browse and download activity) is left as future work.

## Assessing Collaboration Levels

We define two metrics to evaluate the level of collaboration in a community: *content reuse* and *shared user interest*.

- **Content reuse** refers to the percentage of activity in a community that involves existing rather than new content. In a highly dynamic community, where users often add content, harnessing collective action is difficult, if not impossible, as little information accumulates about individual content items.
- **Shared User Interest** characterizes the level of interest overlap among users regarding the content available in the community. This can be inferred from user's preferences regarding existing content, their tagging and/or item consumption (browsing) behavior. Obviously, the higher the shared interest, the higher the chance a user can benefit from information inferred from past activity of other users (e.g., recommender systems, spam filters).

Note that low content reuse implies low shared interest while the converse does not hold; high-content reuse does not imply high shared interest as a single user might reuse the same content. Next, we estimate the level of collaboration in CiteULike and Connotea in light of these two metrics.

## Content Reuse

Content reuse refers to the probability that activity in a community relates to existing (items, tags) rather than to new content.

High content reuse means that, it is more likely that users reuse existing content than add new content to the community. By "reuse" we assume all possible events that refer to an item or a tag. For example, in a collaborative tagging community the same item may be tagged multiple times by the same user (or by multiple users), who can comment on the item, download the item, or simply browse the page without explicitly "interacting" with the content. The higher the content reuse, the more useful information on past activity is to predict future behavior.

In the following, due to the particularities of our traces, we focus only on tag assignments to evaluate content reuse. To this end, we analyze the degree to which tagging activity refers to new versus existing entities, over time, as the communities grow, along three dimensions: items, users and tags. An entity is *new* at time *T*, if there is no report of the entity in the activity trace before *T*. Conversely, a *reused* entity is an item, tag or user which appears in the trace before time *T*. This concept is applied to estimate item and tag reuse, and trace activity to new or existing users.

|  |  | **CiteULike** | **Connotea** |
|---|---|---|---|
| **Reused Items** | Average | 196.13 *(18.43%)* | 26.84 *(7.86%)* |
|  | S.Deviation | 347.30 *(11.74%)* | 31.47 *(6.88%)* |
|  | Median | 72.00 (16.12%) | 19.00 *(6.81%)* |
| **Reused Tags** | Average | 3553.90 *(89.92%)* | 310.19 *(69.70%)* |
|  | S.Deviation | 5888.40 *(8.94%)* | 232.77 *(17.55%)* |
|  | Median | 1314.00 *(92.41%)* | 282.00 *(76.23%)* |
| **Existent users activity** | Average | 102.21 *(79.35%)* | 58.01 *(81.72%)* |
|  | S.Deviation | 77.61 *(11.11%)* | 41.32 *(13.86%)* |
|  | Median | 87.00 *(82.01%)* | 52.00 *(84.61%)* |

**Table 2: A summary of daily item and tag reuse, and user activity in absolute values followed by percetages between brackets.**

In summary we find that, both communities present the following major characteristics: (1) *consistently low levels of item reuse*, (2) *high levels of tag reuse*, and (3) *most activity being generated by existing users*. Table 2 summarizes the content reuse and active user population in CiteULike and Connotea. Due to space constraints and since these observations are similar to both CiteULike and Connotea, we present below only CiteULike data.

Figure 1 presents the percentage of daily activity for new and reused items since the CiteULike community creation. The activity related to new items significantly dominates the level of item reuse, by a factor between four and nine. Simply put, *users are adding new items much faster than they are reusing them*. However, CiteULike exhibits a trend of growing item reuse in the last six months of activity, trend which is not visible in Connotea.

To understand whether the low item reuse ratio we observe is due to a high rate of new users joining the community, we measure the percentage of daily tagging activity performed by new and existent users. These results, reported in Figure 2, show that the largest portion of the activity (about 80%) is generated by existing users, result that holds for both CiteULike and Connotea.

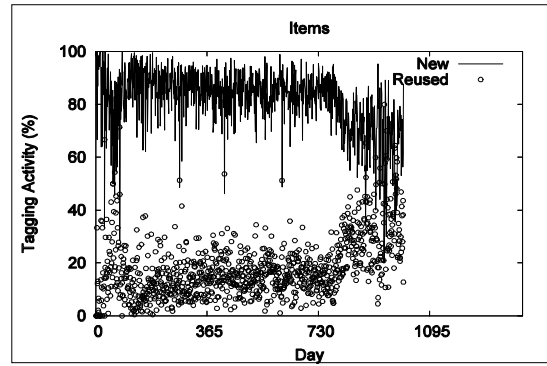

**Figure 1: Item reuse in CiteULike as measured daily. The level of reuse is low, 18.4% on average (points on the bottom).**

We now turn our attention to tag reuse. Figure 3 presents the level of daily tag reuse in CiteULike. Clearly, tags are reused at a higher rate than items. In fact, the activity involving existing tags in CiteULike is eight times larger than new tags, while in Connotea it is three times larger.

Finally, we note that both item and tag reuse as well as the share of activity generated by new users appears to stabilize over time. The "noisy" part (left hand side in our plots) is observed exactly during the initial operation of these communities (Cameron 2007).

**Summary**: Both communities we analyze consistently display low levels of item reuse, high levels of tag reuse, and the majority of the activity being generated by existing users. However: *Do users tag each other's content? If they do so, to what degree do users share the same "opinion" about the same items? What can we say about the characteristics of the structure of shared user-interest?*

## Shared Interest

A second way to characterize user behavior in a collaborative tagging community is to focus on the pair-wise shared interest between users, as reflected by their activities. We seek to understand whether the high level of tag reuse results in users that are tagging overlapping sets of items and/or use overlapping sets of tags.

To this end, this section formalizes the notion of shared interest between a pair of users and presents an evaluation of the level of shared interest in CiteULike (we are still analyzing Connotea dataset). In particular, the analysis of the level of shared interest consists of two parts: first, in this section, the characteristics of the pair-wise interest sharing relation among users; the next section the structure of interest sharing at the community level as displayed by the interest-sharing graph.

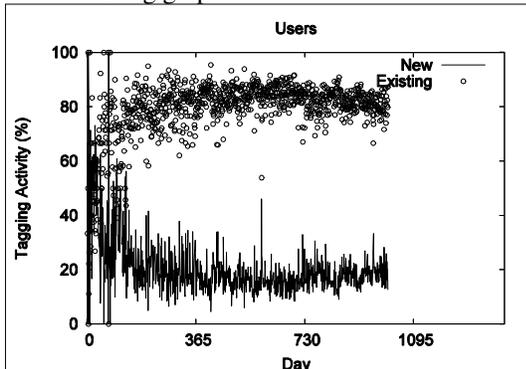

**Figure 2: Daily activity performed by new and existent users in CiteULike. The largest portion of activity is performed by existent users (79.39% on average).**

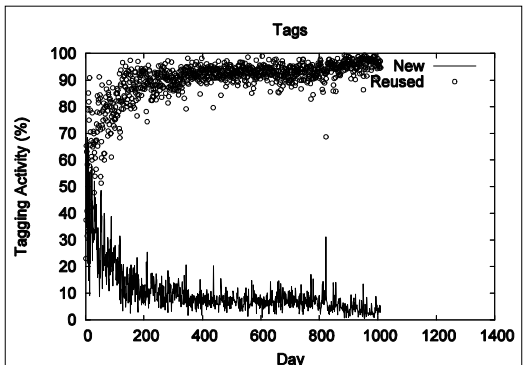

**Figure 3: Percentage of tag reuse as measured daily, in CiteULike. Users tend to significantly reuse tags.**

|  |  |  |
|---|---|---|
| **User-Item** | Average | 0.076119 |
|  | Standard Deviation | 0.166700 |
|  | Median | 0.023256 |
| **User-Tag** | Average | 0.131230 |
|  | Standard Deviation | 0.272550 |
|  | Median | 0.022222 |

**Table 3: Summary of the level of user shared interest in CiteULike according to definitions in Equations 1 and 2.**

*Terminology.* Let a tagging community be represented by a tuple $C:=(U,I,T,A)$, where $U$ represents the set of users, $I$ is the set of items (or library), $T$ is the set of tags (or vocabulary), and $A$ is the set of tag assignments. Specifically, the set of tag assignments is defined as: $A := \{(u, t, p, s) \mid u \in U, t \in T, p \in I, s$ is a timestamp$\}$.

The activity of a user is characterized by the tag assignments she performs. Thus, the activity of user $u_k$ is $u_k := (I_k, T_k, A_k)$, where $I_k \subseteq I$ is the set of items a user $u_k$ posted (or her library), $T_k \subseteq T$ is the set of tags assigned by $u_k$ to her items (or her vocabulary), and $A_k \subseteq A$ is the set of tag assignments performed by $u_k$.

*Metrics for shared interest.* Let us now consider the definition of two functions that quantify the shared interest between two users. The definitions are stated below and formalized by Equations 1 and 2, respectively.

**Definition 1** (*User-Item*)**:** *The* item *interest sharing of two users is the ratio between the size of the intersection and that of the union of their item-sets.*

$$w_I(k, j) = \frac{|I_k \cap I_j|}{|I_k \cup I_j|} \quad (1)$$

**Definition 2 (*User-Tag*):** *The* tag *interest sharing of two users is ratio between the intersection and the union of their vocabularies.*

$$w_T(k, j) = \frac{|T_k \cap T_j|}{|T_k \cup T_j|} \quad (2)$$

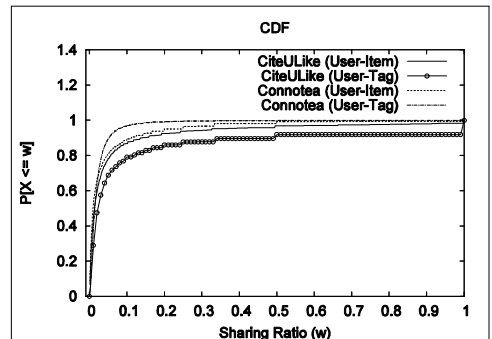

**Figure 4: Cumulative probability distributions of the sharing ratio in CiteULike and Connotea for the User-Item and User-Tag interest sharing graph. More than 85% of pairs of users have a User-Item interest-sharing level lower than 0.05, and about 75% of users has a User-Tag interest-sharing level lower than 0.03 in CiteULike. Connotea has but even lower levels.**

Using the model defined in the previous section, it is possible to assess the level of shared interest in the community by measuring the pair-wise shared interest. First, Table 3 presents the average, standard deviation and median for these item- and tag- interest sharing to give an initial intuition on the overall level of shared interest.

The main lesson is that the average and median of sharing ratios (according to both definitions) are noticeably small. As expected, the sharing ratio considering only items (User-Item in the table) presents the lowest average and median of both definitions. Finally, we note that the coefficient of variation (i.e. ratio between average and standard deviation) is high which prompts us to analyze the interest sharing distribution across all pairs of users.

Figure 4 shows that the pair-wise interest sharing is concentrated: approximately 85% of the pairs have sharing ratios values smaller than 0.05 for the User-Item and User-

Tag definitions. This is a strong indication that shared interest is concentrated: only a small number of user pairs share interest over items and use the same tags (not necessarily on the same items). Furthermore, the level of shared interest presented in Figure 4 suggests that the largest portion of the tag reuse observed in Figure 3 is due to individual user reuse of tags (i.e. the same users repeat the same tag to several items, e.g., to group items in categories), rather than multiple users using the same tag.

To identify any evolution trend of interest-sharing, we evaluate how the levels over time. Figures 5 and 6 present a box-plot of the interest-sharing level measured monthly, according to the User-Item and User-Tag definitions, respectively. The important aspect is that the median and the upper quartile is consistently low overtime for both interest-sharing definitions, which supports the previously observed interest-sharing level when considering the entire data set. Finally, in Figure 5 presents an increase in the User-Item shared interest level, while there is a clear decrease in the User-Tag shared interest level. Although this aspect demands deeper investigation, it suggests that users start reusing content, while *disagreeing* on the tags that best describe the shared content.

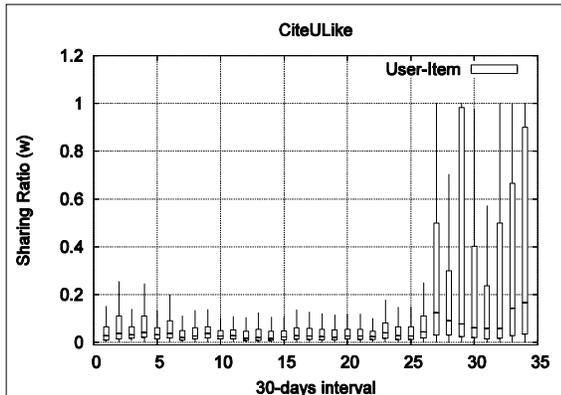

**Figure 5: Levels of Interest-sharing in intervals of 30 days for CiteULike.**

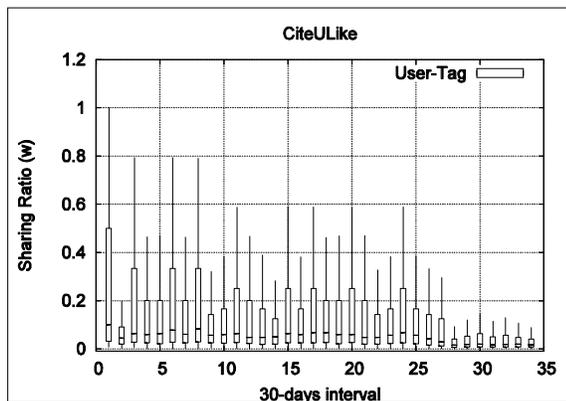

**Figure 6: Interest-sharing evolution (User-Tag) in CiteULike.**

## The Structure of Interest-Sharing

To study the *structure* of interest-sharing in the tagging community, we define a weighted graph where nodes are the users, and a link between two users is labeled with the (item or tag) interest sharing ratio between those two users. This graph formulation is inspired by the data-sharing graph introduced by (Iamnitchi et al. 2004) and we refer to it as the *interest-sharing graph*.

Our assumption is that the topological analysis of the interest-sharing graph will reveal the patterns of interest sharing in the tagging community. The rest of this section summarizes our early results in this direction.

We use thresholds corresponding to the knee of the distributions in Figure 4 to eliminate all the links corresponding to low shared interest in the interest-sharing graph. After this, the resulting has remarkable topological properties:

1. *A large population of isolated* users (zero-degree nodes in the interest-sharing graph). This indicates that there are a large number of users with *unique* preferences. The set of isolated nodes represents around 53% of the population for user-item interest sharing. Nevertheless, the opposite is observed in the user-tag interest sharing, where only 2% of users are isolated.
2. *A significant number of small sub-communities* of interests totally separated from each other. In fact, 15% or of the nodes fall in this category when user-item interest sharing is used, while under the user-tag interest sharing only a negligible percentage is observed in this category.
3. *A dense core,* with an average clustering coefficient of approximately 0.66 for the user-item interest sharing, and 0.17 for the user-tag interest sharing.

These preliminary results suggest that, while the overall levels of collaboration are low, we may still be able to gather social knowledge if we ignore users with isolated interests and focus our analysis on the dense core of users with overlapping interests.

## Discussion

So far, this paper introduced two metrics (*content reuse* and *shared interest level*) to estimate the level of user collaboration in online tagging communities and presented evidence to support our claim that the level of collaboration in tagging communities is lower than generally assumed in the literature. These findings have implications on the ability to harvest social knowledge in these environments. In fact, the low levels of content reuse and shared interest can be mapped to known problems in the domain of recommender systems (i.e., "new items" and sparsity, respectively (Desphande et al. 2004) (Chen et al. 2006) (Adomavicius et al. 2006)).

This section discusses the limitations of the present study, the implications of our findings and, finally, it suggests a tentative roadmap for future research.

**Limits of our study**: There are two important limitations to our study: breadth and type of activity. First, we only study a limited number of tagging communities. We consider, however, that these communities are representative for the larger class of collaborative tagging based on their openness and popularity (tens of thousands of users). Second, the traces allow characterization of these communities and evaluation of collaborative usage only based on explicit content sharing (item posting and tag assignment), but do not allow us to estimate collaboration levels based on content consumption (browsing, for instance). We are in the process of obtaining browsing traces from a smaller tagging community and plan to analyze these traces to better understand the bias we introduce by limiting our analysis to content creation.

The low level of user collaboration we document impacts the design of the websites that support tagging communities. More importantly, low collaboration levels impact the ability to harness the social knowledge produced. We briefly discuss two potential application domains: recommendation systems and reputation systems.

**Content management infrastructures should target individual first and collaborative usage second**. This is a view long held by experts (Grudin 1994) (Golder and Huberman 2007) (Iverson 2007), and our study offers quantitative data to support this view: Collaboration does not always naturally emerge, and the current popularity of existing collaborative tagging sites is a result of their ability to cater to the demands of individual users rather than a direct consequence of their ability to aggregate social knowledge. A number of design recommendations for new collaborative tagging projects result from this observation. Paramount is to design the tagging website such that it caters to the personal information management needs of individual users first, while placing data gathering for social knowledge discovery transparently in the background.

**Recommendation systems** rely on the similarity between users' interests to produce suggestions. However, due to the low levels of content reuse and shared interest, the effectiveness of recommendation systems in the tagging communities we analyze tends to be low. We have prototyped a relatively simple recommendation system based on the interest-sharing graph and have evaluated its efficiency using CiteULike data (Santos-Neto et al. 2007). The success rate of our system is 10-20% (we consider a recommendation successful when the *actual* user activity is included in the set of items or tags that our system *recommends*). We note that the low levels of content reuse displayed by CiteULike highly impact the success rate of the recommendation system: The success rate of the recommendation system is higher than 90% when we restrict our predictions to items that are reused.

This experiment provides preliminary evidence that social knowledge can be efficiently harnessed in online tagging communities provided that the content of communities starts to stabilize.

**Reputation systems:** due to the low levels of content reuse and shared interest, maliciously inserted content (items or tags) is hard to detect in collaborative tagging communities. Link-based ranking algorithms, such as PageRank may be applied over the interest-sharing graph to find authoritative sources for content. However, the large share of users with non-overlapping interests is likely to limit the efficiency of such algorithms, since there is no information that can be extracted to infer the reputation of these users based on the link structure. Additionally, the low level of content reuse implies that, for a large number of items, no reputation data can be inferred as they are recently added. A potential solution that may be worth investigating is to augment the reputation extraction algorithms based on explicit content sharing combined with implicit usage patterns such as browsing histories.